\documentclass{iau_FM}
\usepackage{graphicx}
\usepackage{newtxtext,newtxmath}
\usepackage{amsmath}	

\usepackage{natbib}

\usepackage{ulem}
\usepackage{tabularx}
\usepackage{bm,url}
\usepackage{times}
\usepackage{color}
\usepackage{float}
\usepackage{arydshln}
\usepackage{multirow}
\usepackage{xcolor}
\usepackage[switch]{lineno}

\usepackage{multicol}
\graphicspath{{./fig/}{./png/}}

\def\Rs{R_{s}}
\newcommand{\bl}{Babcock--Leighton}

\newcommand{\Eq}[1]{Eq.~(\ref{#1})}

\def\Rs{R_{s}}
\newcommand{\etasurf}{\eta_{\mathrm{surf}}}
\newcommand{\etaSCZ}{\eta_{\mathrm{SCZ}}}
\newcommand{\etaRZ}{\eta_{\mathrm{RZ}}}
\newcommand{\rBCZ}{r_{\mathrm{BCZ}}}
\newcommand{\rsurf}{r_{\mathrm{surf}}}

\title[Solar and stellar dynamos] 
{Modelling the occurrence of grand minima in sun-like stars using a dynamo model}

\author[V. Vashishth]   
{Vindya Vashishth
}

\affiliation{Department of Physics, Indian Institute of Technology (Banaras Hindu University) Varanasi 221005 India \\ email: {\tt vindyavashishth.rs.phy19@itbhu.ac.in
} \\[\affilskip]
}

\pubyear{2022}
\jname{Astronomy in Focus, Volume 1} 
\editors{Piero Benvenuti, ed.}
\begin{document}

\maketitle

\begin{abstract}
In this work, we have studied the variability and frequency of occurrence of the grand minima using kinematic dynamo models of one solar mass star with different rotation rates and depths of convection zones. We specify the large-scale flows (differential rotations and meridional circulations) from corresponding hydrodynamic models. We include stochastic fluctuations in the Babcock-Leighton source for the poloidal field to produce variable stellar cycles. We observe that the rapidly rotating stars produce highly irregular cycles with strong magnetic fields and rarely produce Maunder-like grand minima, whereas the slowly rotating stars (Sun and longer rotation period) produce smooth cycles of weaker strength and occasional grand minima. In general, the number of the grand minima increases with the decrease of rotation rate. These results can be explained by the fact that with the increase of rotation period, the supercriticality of the dynamo decreases and the dynamo is more prone to produce extended grand minima in this regime.


\end{abstract}

\firstsection 
\section{Introduction}

Like our Sun, many other sun-like stars have magnetic fields and cycles as unveiled by various observations \citep{donati92, Baliu95, wright11, WD16}.
According to these observations, the rotation rate of a star plays an important role in determining its magnetic activity. Rapidly rotating (young) Sun-like stars exhibit a high level of activity with no Maunder-like grand minimum (flat activity) and rarely display smooth regular activity cycles. On the other hand, slowly rotating old stars like the Sun and older have lower activity levels and smooth cycles with occasional grand minima \citep{Skumanich72, R84, Baliu95, Olah16, BoroSaikia18, garg19}.
Recently, \citet{Shah18} observed the decreasing magnetic activity of HD 4915 which might be an indication of Maunder minimum candidate. Later \citet{anna2022} confirmed that HD 166620 is enering into a grand minimum phase. Interestingly, these stars (including Sun) are slow rotators.

In the Sun, differential rotation and helical convection are the two crucial ingredients that govern the dynamo. 
This is due to the fact that the toroidal field is generated through the stretching of the poloidal field by the differential rotation, which is known as the  $\Omega$ effect. 
There is strong evidence that the poloidal field in the Sun is generated through a mechanism, so-called the Babcock--Leighton process \citep{Das10, KO11, Muno13, Priy14, KKV21, KBK22, Mord22}. 
In this process, 
tilted sunspots (more generally bipolar magnetic regions) decay and disperse to produce a poloidal field through turbulent diffusion, meridional flow, and differential rotation.
While the systematic tilt in the BMR is crucial to generate the poloidal field, the scatter around Joy's law tilt \citep[e.g,][]{MNL14, Jha20} produces a variation in the solar cycle \citep{LC17, KM17, KM18}.   

Similar to the Sun, the other sun-like stars also have convection zones(CZs) in their outer layers, it is expected that these stars also support similar dynamo action through which the stellar magnetic cycles are maintained.
Some of the stars (e.g., HD 10476, HD 16160, etc.) have cycles similar to the solar cycle which suggests that a similar dynamo that is operating in our Sun might be working in other sun-like stars \cite[also see][]{jeffers22}.

The motivation of our work is to extract the dependency of the rotation rate of the sun-like stars on its cycle variability and the occurrence of the grand minima using the dynamo model of \cite{KKC14} in which, the regular behavior of the stellar cycle was simulated. As the stellar cycles are irregular, it is natural to explore the irregular features of the stellar cycles using this model.
For this, we shall include stochastic noise to capture the inherent fluctuations in the stellar convection as seen in the form of variations in the flux emergence rates and the tilts of BMRs around Joy’s law. To do so, we shall include the stochastic fluctuations in the \bl\ source for the poloidal field in the dynamo.
In this paper, we have included the primary results from the work. A detailed analysis of the work will be described in the upcoming work (Vashishth et al.2022 (under preparation)).

\section{Model}
In our work, we have used a kinematic mean-field dynamo model by assuming the axisymmetry. This model is based on \cite{KKC14}. Thus, the evolution equation of the poloidal ($\nabla \times [ A(r, \theta) {\bf e}_{\phi}]$) and toroidal ($B (r, \theta) {\bf e}_{\phi}$) fields are followings.

\begin{equation}
\frac{\partial A}{\partial t} + \frac{1}{s}({\bf v_p}.\nabla)(s A)
= \eta \left( \nabla^2 - \frac{1}{s^2} \right) A + S(r, \theta; B),
\end{equation}
\begin{eqnarray}
\frac{\partial B}{\partial t}
+ \frac{1}{r} \left[ \frac{\partial}{\partial r}
(r v_r B) + \frac{\partial}{\partial \theta}(v_{\theta} B) \right]
= \eta \left( \nabla^2 - \frac{1}{s^2} \right) B
+ s({\bf B}_p.{\bf \nabla})\Omega + \frac{1}{r}\frac{d\eta}{dr}\frac{\partial{(rB)}}{\partial{r}},
\end{eqnarray}

where $s = r \sin \theta$, ${\bf v_p} = v_r {\bf \hat{ r}} + v_\theta {\bf \hat{ \theta}}$ is the meridional flow, and the $\Omega$ is the angular velocity whose profile is obtained from the mean-field hydrodynamic model of \cite{KO11}, $\eta$ is the turbulent magnetic diffusivity which is written as the function of $r$ alone and take the following form,
\begin{eqnarray}
\eta(r) = \etaRZ + \frac{\etaSCZ}{2}\left[1 + \mathrm{erf} \left(\frac{r - \rBCZ}
{d_t}\right) \right]
+\frac{\etasurf}{2}\left[1 + \mathrm{erf} \left(\frac{r - \rsurf}
{d_2}\right) \right]
\label{eq:eta}
\end{eqnarray}\\
with $\rBCZ=0.7 \Rs$ ($\Rs$ being the stellar radius), $d_t=0.015 \Rs$, $d_2=0.05 \Rs$, $\rsurf = 0.95 \Rs$,
$\etaRZ = 5 \times 10^8$ cm$^2$~s$^{-1}$, $\etaSCZ = 5 \times 10^{10}$ cm$^2$~s$^{-1}$, and
$\etasurf = 2\times10^{12}$ cm$^2$ s$^{-1}$.
$S$ is the source for the 
poloidal field and its parameterised form is written as,
\begin{equation}
 S(r, \theta; B) = \frac{\alpha_{\rm BL}(r,\theta)}{1 + \left( \overline{B} (r_t,\theta)/B_0 \right)^2} \overline{B}(r_t,\theta),
\label{source}
\end{equation}

where $\overline{B}(r_t,\theta)$ is the toroidal field at latitude $\theta$
averaged over the whole tachocline from $r = 0.685 \Rs$ to $r=0.715 \Rs$, and  
$\alpha_{\rm BL}$ is the parameter for Babcock--Leighton process and is written as,
\begin{equation}
\alpha_{\rm BL}(r,\theta)=\frac{\alpha_0}{4}\left[1+\mathrm{erf}\left(\frac{r-r_4}{d_4}\right)\right]\left[1-\mathrm{erf}\left(\frac{r-r_5}{d_5}\right)\right]\times \sin\theta\cos\theta
\label{alpha}
\end{equation}
where, $r_4=0.95 \Rs$, $r_5= \Rs$, $d_4=0.05 \Rs$, $d_5=0.01 \Rs$.
$\alpha_0$ is the measure of the strength of the \bl\ process which is expressed as the dependence on the rotation in the following way,
\begin{eqnarray}
 \alpha_0 = \alpha_{0,s} \frac{T_s}{T},
 \label{eq:alpha0I}
\end{eqnarray}
where $\alpha_{0,s}$ is the value of $\alpha_0$
for the solar case, which is taken as $0.7$ cm~s$^{-1}$ and $T_s$ and $T$ are the rotation period of Sun and the star, respectively.

Since there are some randomnesses in the \bl\ process,
we include fluctuations in the $\alpha$ appearing in \Eq{eq:alpha0I} as, $\alpha_{0,s} = \alpha_{0,s} r, $  
where $r$ is the Gaussian random number with mean and standard deviation ($\sigma$) unity. We keep the value of $\sigma$ the same for all the stars.
 In our model, the value of $\alpha_0$ is updated randomly after a certain time which we take to be one month.

\begin{figure}
\centering
\includegraphics[width=1.05\columnwidth]{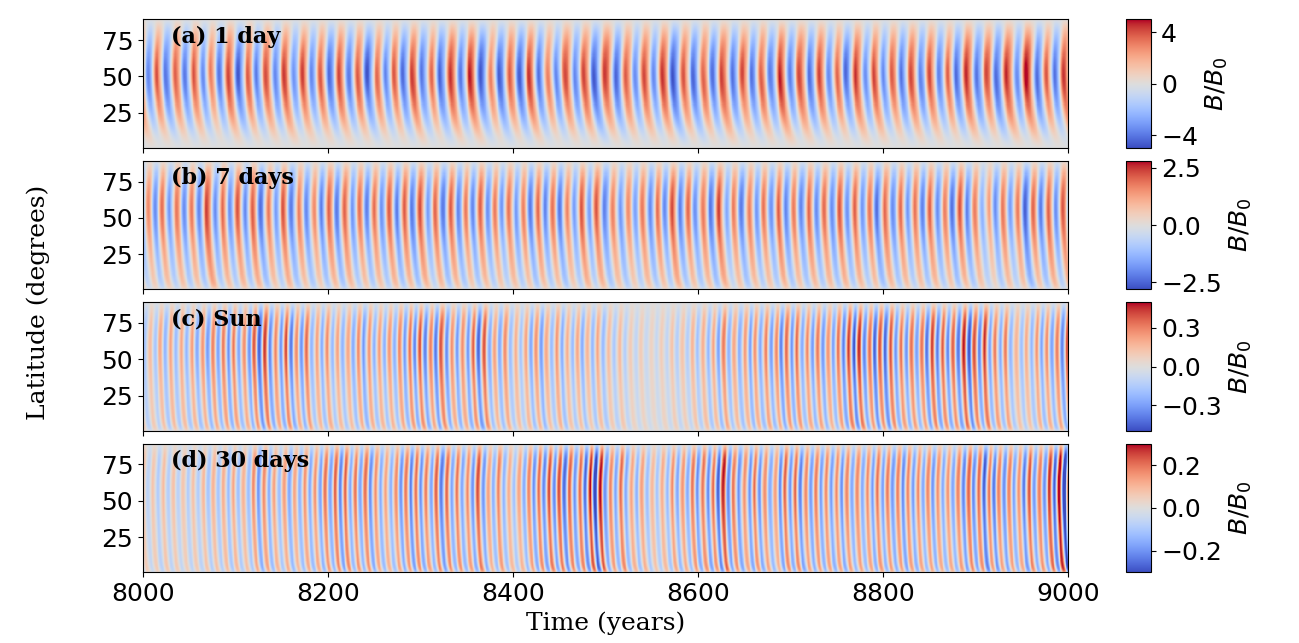} 
 \caption{Time–latitude distribution of the toroidal field at the base of the CZ from different stars.}
   \label{fig1}
\end{figure}

\section{Results}
The simulations were done for $M_\odot$ mass stars having rotation periods of 1, 3, 7, 10, 15, 20, 25.30 (Sun), and 30 days, respectively.
Here we discuss the various aspects of magnetic cycles obtained from all the stars.
Firstly, we find the regular polarity reversal from the time-latitude plot of the toroidal field at the base of the convection zone for all the rotation periods as shown in Fig.\,\ref{fig1}. This figure also depicts a weak equatorward propagation of the field at lower latitudes which is due to the transport of the field by the equatorward meridional circulation.

One obvious feature in these simulations is that in fast-rotating stars, the magnetic field becomes strong. 
This is because the strength of $\alpha$ increases with the rotation rate of the star (the shear however remains more or less unchanged in different stars). 
If a star rotates faster, the tilt of the BMR associated with the \bl\ $\alpha$ is expected to increase. Therefore, with the age, as the rotation rate decreases, the dynamo process becomes weaker and the dynamo number also decreases. This implies that the rapidly rotating stars are likely to have stronger magnetic cycles.
This result is in agreement with the \cite{KKC14} and the observations \citep{Noyes84a, wright11}. 

These time-latitude plots also give a hint about the variability observed in different stars. 
Slowly rotating stars seem to produce more long-term modulation in their cycles, including extended episodes of a weaker magnetic field. 
In contrast, fast rotators generate less long-term modulation.

The root cause for such behavior is that the slowly rotating stars have a small dynamo number. Due to this, if the magnetic cycle gets weaker sometimes, then it would take a long time to grow the field. Therefore, we see a long-term modulation in the slowly rotating stars. But for the fast rotators, the cycle gets stronger much more quickly after getting into the weak phase. This is easy to understand, as, in fast rotators, the dynamo number is high so the growth rate is very high. This trend is also explained in \cite{KKV21, Vindya21}. The results are in accordance with the observations as well \citep{Baliu95}. 

\begin{figure}
\centering
 \includegraphics[width=1.01\columnwidth]{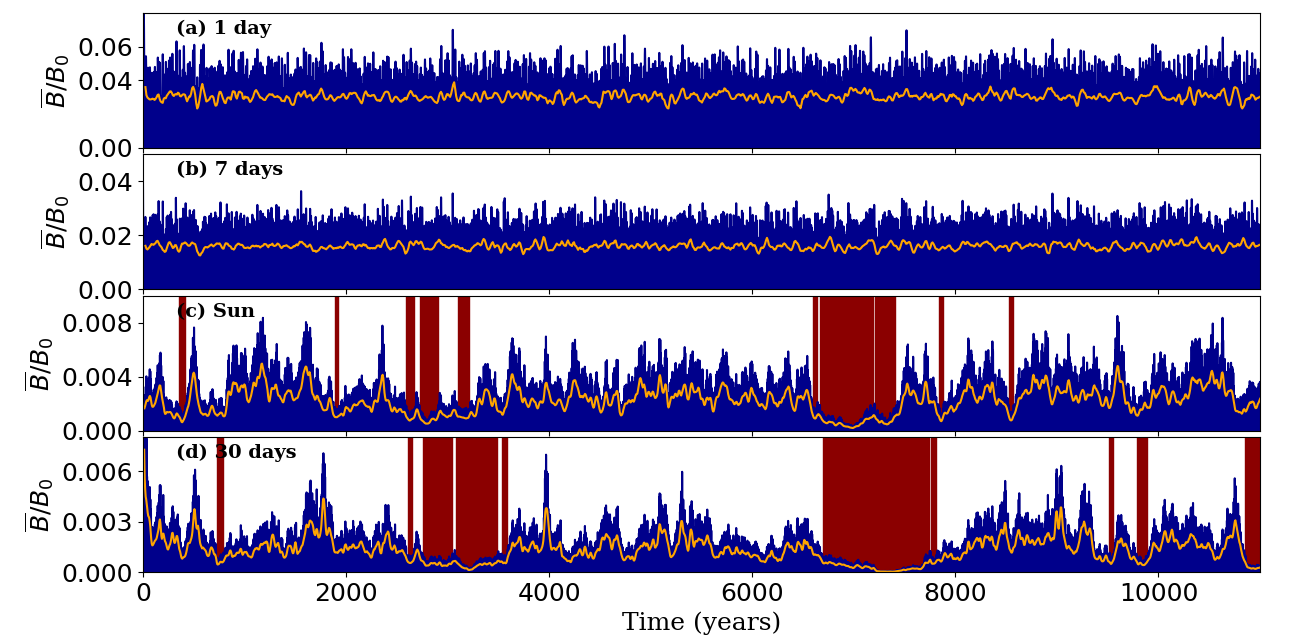} 
 \caption{Time–series of the toroidal field at the base of the CZ for different stars along with their smoothed curves (orange) and grand minima (darkred).}
   \label{fig2}
\end{figure}

We then evaluated the number of grand minima observed in each case with the help of a time-series plot of the toroidal field at the base of CZ from simulation for 11,000 years as shown in Fig.\,\ref{fig2}. We infer that in all the models, the number of grand minima observed shows an increasing trend with the rotation period. We saw that the rapidly rotating stars hardly produce any grand minima whereas the slowly rotating stars produce some grand minima and also as the rotation period increases, the number of grand minima is seen to increase. This is because, with the increase of rotation period, the supercriticality of the dynamo decreases, and the dynamo is more prone to produce extended grand minima in this regime.
This result is as per \cite{Vindya21} where we observed that as the supercriticality increases (i.e. as the rotation period decreases), the frequency of occurrence of grand minima decreases.
This is also supported by the observational fact that the detected grand/Maunder minima candidates are the slow rotators.

\section{Conclusions} 
Based on the kinematic dynamo simulations of one solar mass stars at different rotation rates with stochastically forced \bl\ source, we make the following conclusions.
In slowly rotating stars, the cycles are smooth and show long-term variation with occasional grand minima. Whereas for rapidly rotating stars, the magnetic field is strong, cycles are more irregular, and no grand minima are detected. The number of grand minima increases with the decrease in the rotation rate of the star.

\section{Acknowledgements}
I thank Leonid Kitchatinov for kindly providing the data of large-scale flows of different stars. I acknowledge the travel grant awarded by IAUGA 2022 to attend the in-person conference.
Financial support from the DST through INSPIRE fellowship is also acknowledged. 
\\

\end{document}